\documentclass[%
 aip,
 amsmath,amssymb,
 reprint,%
]{revtex4-1}

\usepackage{soul}       
\setstcolor{red}        

\usepackage{graphicx}
\usepackage{dcolumn}
\usepackage{bm}
\usepackage{xcolor}
\usepackage[utf8]{inputenc}
\usepackage[T1]{fontenc}
\usepackage{mathptmx}
\usepackage{etoolbox}
\makeatletter
\def\@email#1#2{%
 \endgroup
 \patchcmd{\titleblock@produce}
  {\frontmatter@RRAPformat}
  {\frontmatter@RRAPformat{\produce@RRAP{*#1\href{mailto:#2}{#2}}}\frontmatter@RRAPformat}
  {}{}
}%
\makeatother

\makeatletter
\def\@email#1#2{%
 \endgroup
 \patchcmd{\titleblock@produce}
  {\frontmatter@RRAPformat}
  {\frontmatter@RRAPformat{\produce@RRAP{*#1\href{mailto:#2}{#2}}}\frontmatter@RRAPformat}
  {}{}
}%
\makeatother

\makeatletter
\renewcommand\section{\@startsection {section}{1}{\z@}%
  {-1.5ex \@plus -1ex \@minus -.2ex}%
  {0.5ex \@plus.2ex}%
  {\normalfont\normalsize\bfseries\raggedright}}
\renewcommand\subsection{\@startsection {subsection}{2}{\z@}%
  {-1.25ex\@plus -1ex \@minus -.2ex}%
  {0.3ex \@plus .2ex}%
  {\normalfont\normalsize\bfseries\raggedright}}
\makeatother

\begin{document}
\preprint{AIP/123-QED}


\author{Giulia Meucci\textsuperscript{*,}}
\affiliation{Department of Energy Conversion and Storage, Technical University of Denmark, 2800 Kgs.Lyngby, Denmark} \email{giuliam@dtu.dk}

\author{Dags Ol{\v{s}}teins}
\affiliation{Department of Energy Conversion and Storage, Technical University of Denmark, 2800 Kgs.Lyngby, Denmark}

\author{Damon J. Carrad}
\affiliation{Department of Energy Conversion and Storage, Technical University of Denmark, 2800 Kgs.Lyngby, Denmark}



\author{Gunjan Nagda}
\affiliation{Department of Energy Conversion and Storage, Technical University of Denmark, 2800 Kgs.Lyngby, Denmark}

\author{Daria V. Beznasyuk}
\affiliation{Department of Energy Conversion and Storage, Technical University of Denmark, 2800 Kgs.Lyngby, Denmark}

\author{Christian E. N. Petersen}
\affiliation{Department of Energy Conversion and Storage, Technical University of Denmark, 2800 Kgs.Lyngby, Denmark}

\author{Sara Martí-Sánchez}
\affiliation{Catalan Institute of Nanoscience and Nanotechnology (ICN2), CSIC and BIST, Campus UAB, \\ Bellaterra,  Barcelona, Catalonia, Spain}

\author{Jordi Arbiol}
\affiliation{Catalan Institute of Nanoscience and Nanotechnology (ICN2), CSIC and BIST, Campus UAB, \\ Bellaterra,  Barcelona, Catalonia, Spain}\affiliation{ICREA, Passeig de Lluís Companys 23, 08010 Barcelona, Catalonia, Spain}

\author{Thomas Sand Jespersen}
\affiliation{Department of Energy Conversion and Storage, Technical University of Denmark, 2800 Kgs.Lyngby, Denmark}\affiliation{Center For Quantum Devices, Niels Bohr Institute, University of Copenhagen, 2100 Copenhagen, Denmark} 

\date{\today}

\title{Cryogenic performance of field-effect transistors and amplifiers based on selective area grown InAs nanowires}


\begin{abstract}
Indium-Arsenide (InAs) nanowire field-effect transistors (NWFETs) are promising platforms for high-speed, low-power nanoelectronics operating at cryogenic conditions, relevant for quantum information processing.  We  use selective area growth (SAG) of nanowires to realize scalable and planar nanowire device geometries  that are compatible with standard semiconductor processing techniques. NWFETs are fabricated and their low temperature characteristics - including $I_\mathrm{ON}/I_\mathrm{OFF}$ ratios, threshold voltages, sub-threshold slope, interfacial trap density, hysteresis, and mobility - are characterized. The NWFETs operate effectively in integrated circuitry relying on saturation-mode operation. In sub-threshold applications such as amplifiers, we find 
bandwidths exceeding our cryostat wiring, but the gate hysteresis presents challenges for precise tuning of the amplifier operating point. We discuss the role of crystal imperfections and fabrication processes on the transistor characteristics and propose strategies for further improvements. 
\end{abstract}

\maketitle


The prospects of large-scale electronic quantum circuits motivate the continuous search for optimal, scalable, and reproducible material platforms for mesoscopic devices. Selective area growth (SAG) of semiconductor nanowires\cite{friedl:2018, krizek2018field,vaitiekenas:2018} offers an interesting approach by adding scalability to the nanowire platform, which has already proven successful at the single-device prototype level, both as active elements in high-performance field-effect transistors (FETs) at cryogenic conditions\cite{delalamo:2011, zhang:2015, dey:2012, ford:2009, sasaki:2013, dayeh:2009t, storm:2012, nilsson2010temperature, dhara2011facile, jiang2022enhancing, froberg2008heterostructure, Blekker2010high} and for advanced quantum devices\cite{hofstetter:2009,sand-jespersen:2008,larsen:2015, nadj-perge:2010, desplanque2014influence}.\\
\indent Nanowires are traditionally grown by the vapor-liquid-solid (VLS) mechanism \cite{wagner1964vapor} which results in  out-of-plane growth. While such structures are compatible with vertical device architectures\cite{tomioka2012iii, bryllert2006vertical, bryllert2006vertical2}, they pose challenges for integration with standard semiconductor processing, particularly in the context of advanced circuitry for cryogenic and quantum applications which often require multiple electrostatic side-gates along the channel and/or integration of locally coupled charge sensors.  Although progress has been made toward scalable transfer and alignment of VLS-grown nanowires\cite{jia2019nanowire, freer:2010a,yao:2013,fu:2018a}, existing methods often rely on elaborate and partly manual procedures that challenge reproducibility and integration of complex large-scale nanowire circuits. In contrast, SAG allows lithographic control of in-plane growth of single nanowires and nanowire networks and thus opens a route towards scalable nanoelectronics \cite{olsteins:2023, krizek2018field, friedl:2018, friedl2020remote,OphetVeld2020Mar, Sodergren2022May}.
\\
\indent Unlike VLS-grown nanowires, planar selective area grown nanowires are epitaxially clamped to the substrate along their length, and crystal growth occurs over a much larger area. This makes the basic growth mechanism qualitatively different\cite{Beznasyuk2024}, increases the susceptibility to misfit dislocations and material intermixing\cite{beznasyuk:2022, krizek2018field}, and also requires device fabrication directly on the growth substrate. These factors may potentially compromise the electrical performance of selective area grown nanowires and devices.  Here we present fabrication, measurements, and detailed analysis of InAs selective area grown NWFETs and explore their performance in amplifier circuits at cryogenic conditions relevant for applications in integrated control electronics for quantum circuits. 
\begin{figure}[hbt]
\centering
    \includegraphics[width = \linewidth]{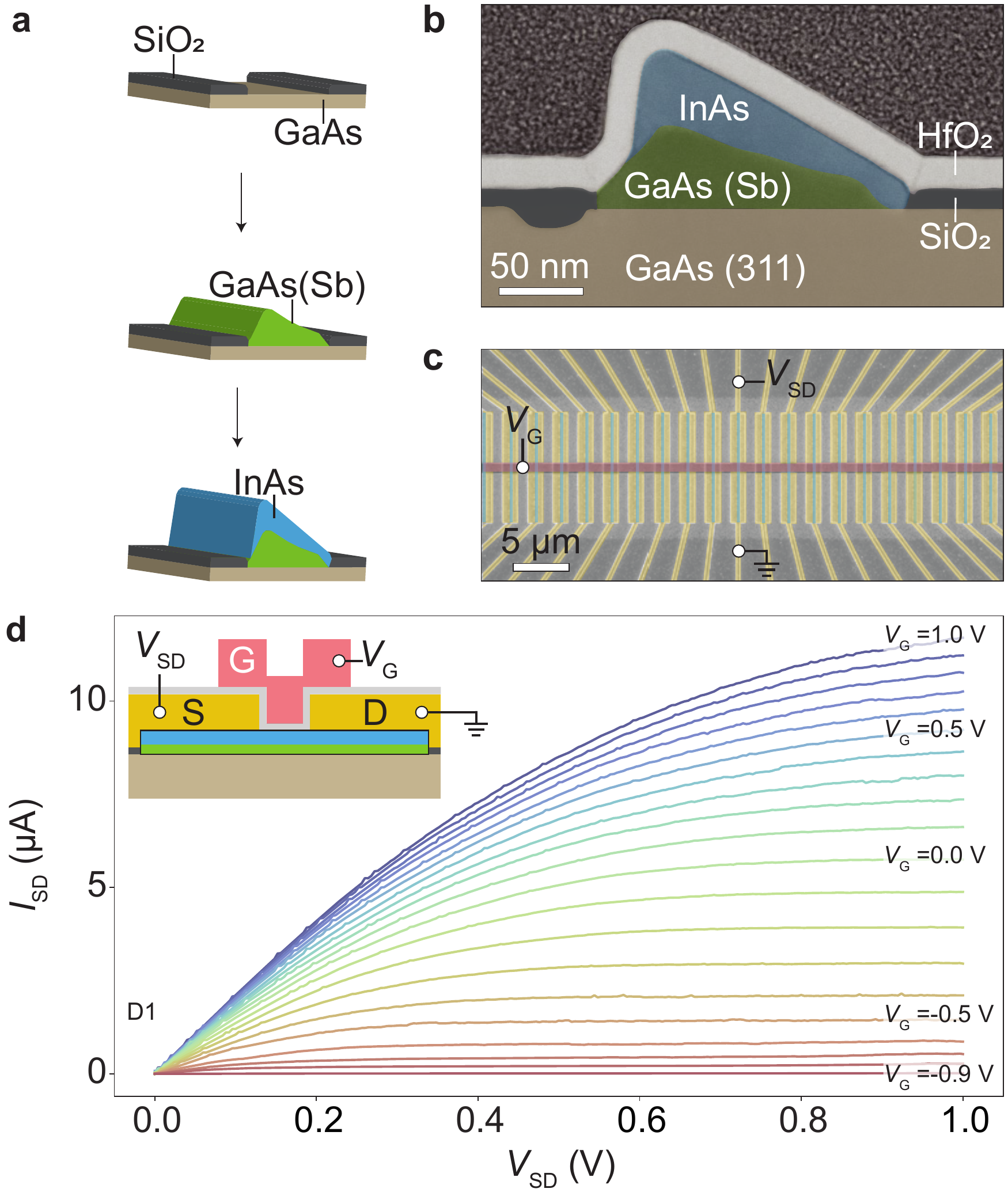}
    \caption{\textbf{a} Schematic illustration of SAG of InAs nanowires. \textbf{b} False-color HAADF STEM micrograph showing the cross-section of a typical nanowire. \textbf{c} False-color SEM micrograph of planar selective area grown nanowire transistors. Each nanowire has individual $\mathrm{Ti}/\mathrm{Au}$ source-drain contacts (yellow) and all devices share a common top gate (pink). \textbf{d} Output characteristics of device D1 at $9$ K with $V_\text{G}$ ranging from $-0.9 \, \mathrm V$ (red) to $1.0 \, \mathrm V$ (blue), in steps of $0.1 \, \mathrm V$. The inset shows a schematic of the device layout. 
    } 
    \label{Fig1}
\end{figure}
\\
\indent 
In-plane SAG of InAs nanowires was performed by molecular beam epitaxy (MBE) using insulating GaAs $(311)$ substrates which were first covered with $10 \, \mathrm{nm}$ of $\mathrm{SiO}_2$ by plasma-enhanced chemical vapor deposition. Subsequently the GaAs surface was re-exposed in $0.15 \times 10 \, \mu \mathrm{m}$ nanowire-shaped regions along the [0$\bar{1}$1] direction using electron beam lithography and CF$_4$ plasma etching. The surface was prepared for the InAs growth by annealing at $610 ^\circ$C to remove native oxides followed by selective growth of a $\mathrm{GaAs(Sb)}$ buffer layer\cite{olsteins:2023} (see Fig.\ \ref{Fig1}a). A cross-section high-angle annular dark-field scanning transmission micrograph (HAADF STEM) of a typical nanowire is shown in Fig.\ 1b. The combination of $(311)$ growth substrate and [0$\bar{1}$1] in-plane orientation yields an asymmetric triangular nanowire cross-section bound by \{111\}A facets. Details of growth and structural analysis is reported elsewhere\cite{olsteins:2024,olsteins:2023}.   Due to the band-bending at the InAs/GaAs interface the InAs layer constitute the electrical transport channel of the structure, as previously confirmed by electrostatic simulations \cite{beznasyuk:2022}. FETs with Ti/Au ohmic contacts and a source-drain separation of $1 \, \mu \mathrm m$ were fabricated directly on the growth substrate by electron-beam evaporation, after removal of the native InAs oxide by in situ radio-frequency (RF) ion milling. A Ti/Au  top gate, separated from the nanowire by 15 nm of $\mathrm{HfO}_2$ grown by atomic layer deposition, tunes the carrier density of the devices.   Typical devices are shown in Fig.\ 1c. The devices were incorporated into an on-chip multiplexer circuit, as described in Ref.\ \citenum{olsteins:2023}, and here we analyze the details of individual FET devices. The NWFETs were characterized by measuring the dependence of the source-drain current, $I_\text{SD}$, on the source-drain voltage, $V_\text{SD}$ (output characteristics), and on the gate voltage, $V_\text{G}$ (transfer characteristics), at temperatures down to $T=14$ mK, in a dilution cryostat. In total, eight nominally identical devices (D1-D8) were studied, with different measurements performed for different devices. An overview of the characterization parameters for each device can be found in supplementary section S1. Device D8 showed effects of quantum interference and ballistic transport at low temperature and low bias (supplementary section S3), and was excluded from the analysis of FET parameters. The main text shows representative data and additional results are presented in the supplementary material.\\
\indent Figure \ref{Fig1}d shows the output characteristics of D1, measured at $9$ K, for $V_\text{G}$ ranging between $-0.9$ and $1 \, \mathrm V$. An ohmic regime is observed at low $V_\mathrm{SD}$, followed by a saturation regime at higher bias,  consistent with previous results for InAs NWFETs at room temperature  \cite{sasaki:2013, ford:2009, dey:2012, nilsson2010temperature, dhara2011facile, froberg2008heterostructure, Blekker2010high}, demonstrating that the selective area grown device acts as a  conventional  FET even at cryogenic temperatures. A zoom-in of the ohmic regime is provided in supplementary section S2.
\\
\indent Figures  \ref{Fig2}a,b show the transfer characteristics of D4, measured at $14$ mK, for  values of  $V_\text{SD}$ between $20$ and $100$ mV, in linear and logarithmic scale, respectively. Similar measurements  for other devices are presented in supplementary section S3. 

Due to limited measurement sensitivity in the sub-nanoampere range the current saturates at low $V_\mathrm G$  to a bias-independent value of $\sim 0.35 \, \mathrm{nA}$, which serves as an upper bound on the off-state current, $I_\mathrm{OFF}$, while the  ON current, $I_\text{ON}$, is  defined as the current at the highest $V_\text{G}$. Based on the $V_\mathrm{SD} = 100\, \mathrm{mV}$ transfer characteristics the corresponding lower bound on the $I_\mathrm{ON}/I_\mathrm{OFF}$ ratio is then $5 \times 10^3$. This value is lower, and thus consistent, compared to those reported for VLS InAs FETs\cite{storm:2012} and further experiments are needed to determine and quantify any significant differences between the two growth methods.
\begin{figure}[hbt]
  \centering
    \includegraphics[width = \linewidth]{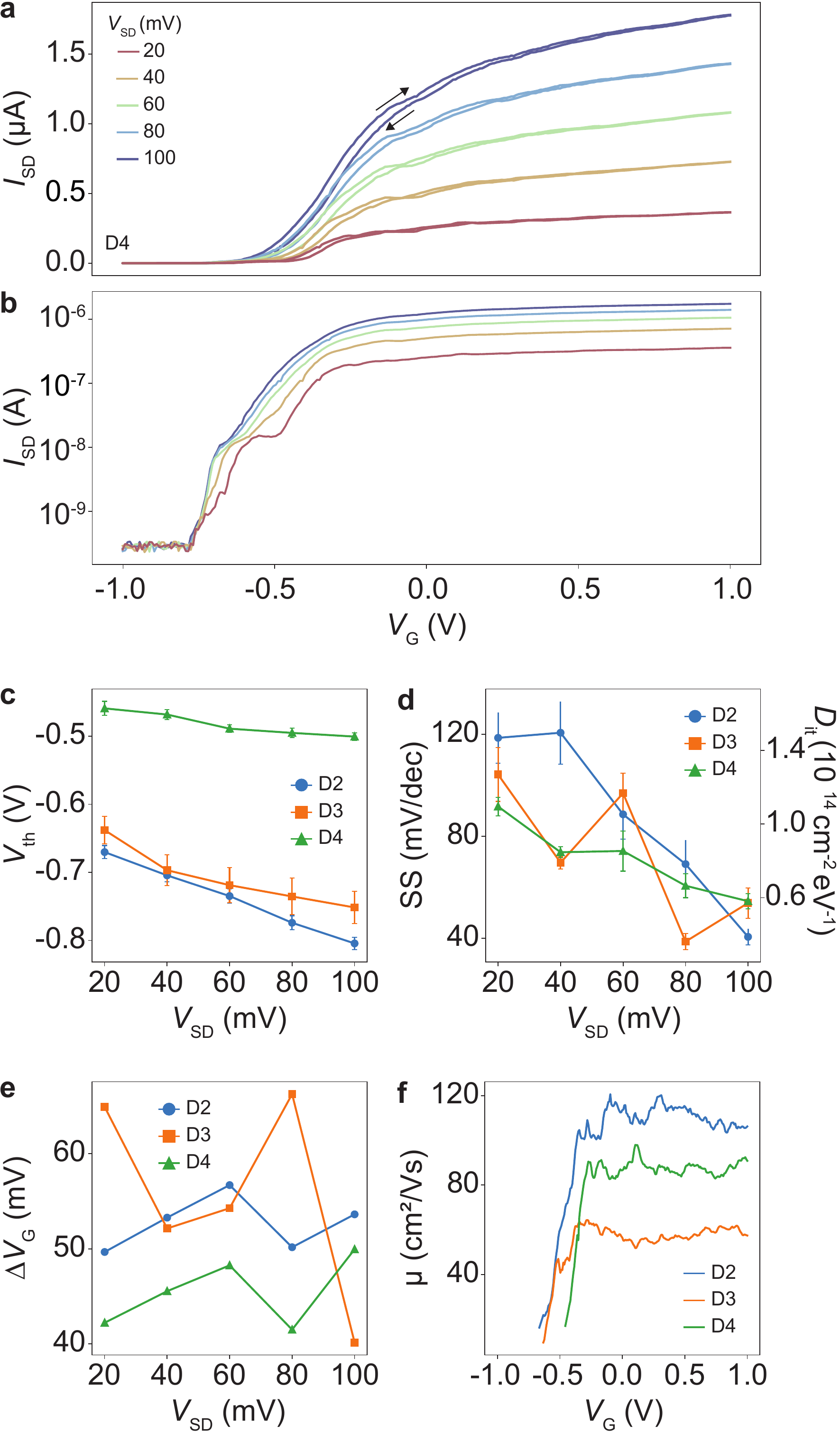}
    \caption{\textbf{a} Transfer characteristics of D4 measured at 14 mK for $V_\mathrm{SD}$ ranging from $20 \, \mathrm{mV}$ (red) to $100 \, \mathrm{mV}$ (blue) in steps of 20 mV. \textbf{b} Curves with positive sweep direcition from panel \textbf{a} shown on logarithmic scale. \textbf{c, d, e, f} Bias dependence of the lower bound of the $I_\mathrm{ON}/I_\mathrm{OFF}$ ratio (\textbf{c}), threshold voltage (\textbf{d}),  inverse sub-threshold slope and associated density of interfacial charge traps (\textbf{e}), and gate hysteresis (\textbf{f}).    \textbf{g} $V_\text{G}$-dependence of the field effect mobility at $V_\mathrm{SD}=20 \, \mathrm{mV}$. 
    } 
    \label{Fig2}
  \hfill
\end{figure}
\\
\indent To estimate the threshold voltage, $V_\text{th}$, we  follow the method outlined in Ref.\ \citenum{gatedependentmobility} (see supplementary section S4). The results are shown in Fig.\ \ref{Fig2}c as a function of $V_\text{SD}$. For all devices, a negative $V_\text{th}$ is observed, which shows that the devices act as $n$-type depletion mode FETs, as expected since nominally undoped InAs forms an electron accumulation layer at the surface, shifting the Fermi level above the conduction band edge \cite{olsson1996charge}. Moreover, $V_\text{th}$ decreases with $V_\text{SD}$, indicating significant drain-induced barrier lowering (DIBL) of $1.7 \pm 0.2$, $1.3 \pm 0.4$, and $0.5 \pm 0.1$ V/V for devices D2, D3, and D4, respectively. These values are consistent with those reported for InAs NWFET with similar architecture and aspect ratios \cite{yang:2018}. Lower values may be obtained through optimized device designs, for example by using a thinner gate-dielectric or SAG substrate symmetries and nanowire orientations leading to nanowire cross-sections more susceptible to gating \cite{Lee2019Aug, nagda2023effect}. We note that rectangular cross-sections close to high-performance FinFET geometries can be achieved using SAG\cite{Lee2019Aug}, which could help reduce the short-channel effects observed here (e.g. DIBL)\cite{delalamo:2011}, although planar SAG structures remain incompatible with the gate-all-around architectures developed for for high performance FETs based on out-of-plane nanowire growth\cite{zhang:2015, tomioka2012iii, tomioka2015selective}.  \\
\indent As shown in Fig.\ \ref{Fig2}b, $I_\text{SD}$ exhibits an exponential increase with $V_\text{G}$ for $V_\mathrm{G} < V_\mathrm{th}$, as expected for FETs. To analyze this sub-threshold regime the transfer characteristics were fitted to the standard expression $I_{\text{SD}} = I_{\text{OFF}} \times 10^{(V_{\text{G}} - V_{\text{t}})/\text{SS}}$, where $V_\text{t}$ and the inverse sub-threshold slope, SS, are fitting parameters \cite{sze2021physics}. We take the fitting range extending from the highest $V_\text{G}$ where $I_\mathrm{SD} < I_\text{OFF} + 10^{-11} \mathrm A$ to the lowest $V_\mathrm{G}$ where  $I_\mathrm{SD} > I_\text{OFF} \times 10^{2}$. This convention allows  consistent examination of the sub-threshold region for the different devices. Figure \ref{Fig2}d shows the resulting SS for different devices and $V_\text{SD}$. SS falls within the range $40-120$ mV/decade similar to reported values for top-gated InAs NWFETs\cite{vasen:2016,zhang:2015d,fu:2014,li:2014a,yang:2018,jiang2022enhancing}. In the cryogenic limit we expect $\text{SS} \approx W_\mathrm t\ln(10) (1+e^2 D_\mathrm{it} A_\mathrm G/C_\mathrm G)/e$, where $W_\mathrm t \approx 5.7 \, \mathrm{meV}$ is the characteristic decay of the conduction band tail in the bandgap \cite{beckers2019theoretical,sarangapani2019band}, $C_\mathrm G$ is the gate capacitance, $A_\mathrm G$ is the gate area, and $D_\text{it}$ is the density of charge traps at the InAs/$\mathrm{HfO}_2$ interface, reducing the effectiveness of the gate.
The devices here have $A_\mathrm G = 180 \, \mathrm{nm} \times 1\mu \mathrm m$, and $C_\text{G}=5.3 \, \mathrm{fF}$ was determined by numerical simulations \cite{olsteins:2023}. With these values we find $D_\mathrm{it} \sim 10^{14}$  cm$^{-2}$ eV$^{-1}$ (Fig.\, \ref{Fig2}d), one order of magnitude lower than values reported for back-gated InAs VLS NWFETs\cite{li:2018a} while an order higher than reported for optimized top gated devices\cite{jiang2022enhancing, zhang:2015}. This suggests that our selective area grown devices can be further improved by systematic optimization of the oxide/nanowire interface \cite{riel:2014}.\\
\indent The presence of interfacial charge traps is consistent with the up/down $V_\mathrm G$ sweep hysteresis of the transfer curves \cite{olsteins:2024}, observed in Fig.\ \ref{Fig2}a. To quantify the hysteresis, we estimate the voltage shift, $\Delta V_\mathrm G$, between the up and  down sweep curves (see supplementary section S5 for details). 
 As shown in Fig.\ \ref{Fig2}e $\Delta V_\mathrm G$ ranges from $40 $ to $60 \, \mathrm{mV}$  which is significantly lower than the typical width of the transfer characteristics, $\sim 150-600 \, \mathrm{mV}$, estimated from the width of a Lorentzian fit to the peaked trans-conductance $\mathrm d I_\mathrm{SD}/\mathrm d V_\mathrm G$. This enables the use of the NWFETs in circuitry requiring switching and saturation, such as multiplexers\cite{olsteins:2023} or other key components in integrated control electronics.
The role of hysteresis in sub-threshold applications, like amplifiers, is discussed below.  \\
\indent To complete the electrical characterization, the mobility, $\mu$, of the NWFETs is assessed based on the $V_\text{SD}$=20 mV data following the procedure outlined in Ref.\ \citenum{gatedependentmobility} (see supplementary section S4).  As shown in Fig.\ \ref{Fig2}f $\mu$ remains almost constant for $V_\text{G}>V_\text{th}$, with values in the range 60-120 cm$^2$/Vs.
The $\mu$ values are lower compared to other reported InAs nanowires, grown using both VLS  \cite{sasaki:2013, ford:2009, storm:2012, dayeh:2009t, dey:2012, wang2019shape, dhara2011facile, Blekker2010high} and SAG \cite{krizek2018field, friedl2020remote, tomioka2015selective}. We attribute this difference to the growth limitations discussed above. We expect that substantial enhancements of $\mu$ can be obtained, for example, by including an InGaAs buffer layer in the selective area grown nanowire heterostructure, as demonstrated  in Ref.\ \citenum{beznasyuk:2022}.  \\
\begin{figure}[hbt]
  \centering
    \includegraphics[width = \linewidth]{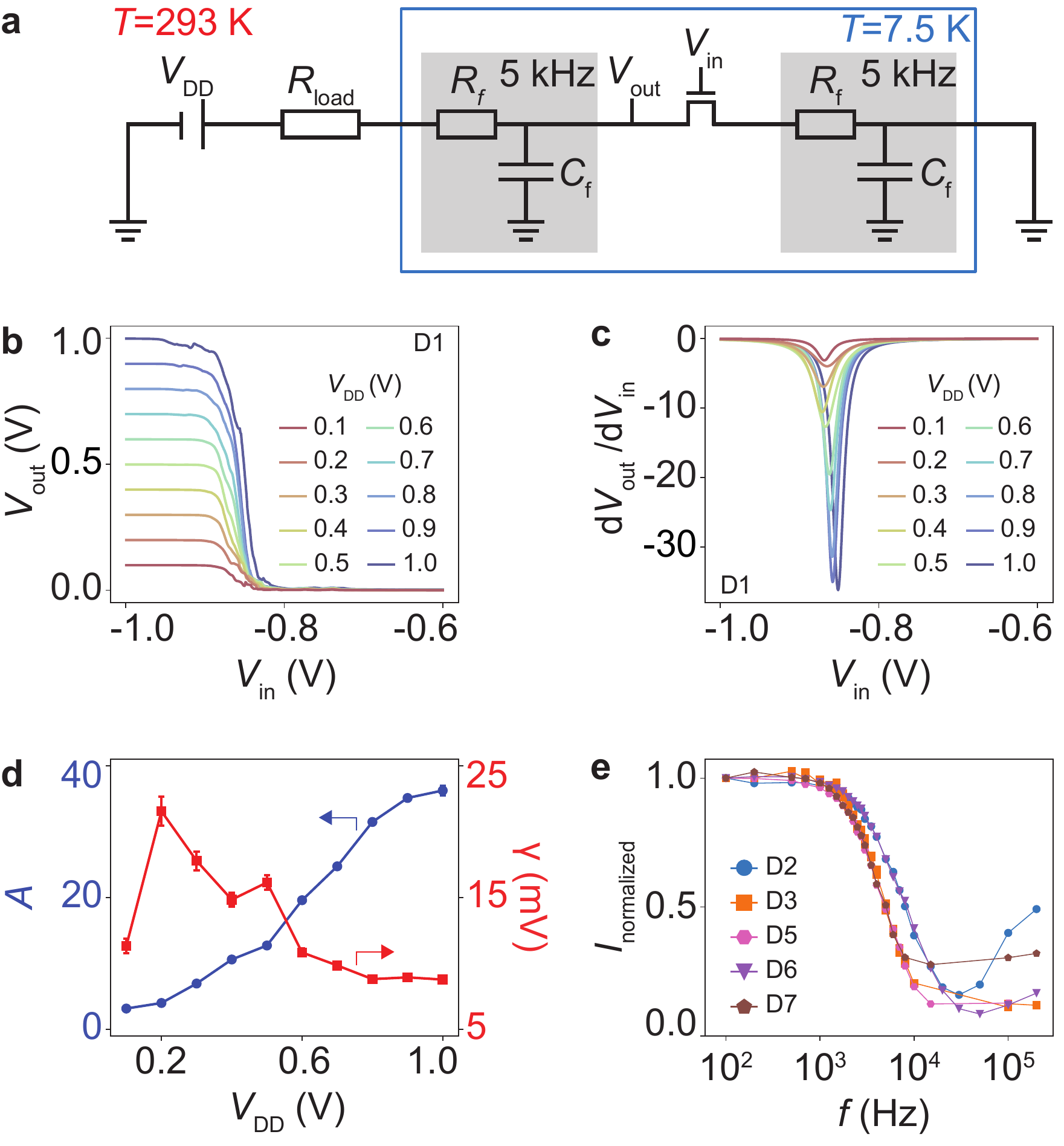}
    \caption{\textbf{a} Common source amplifier circuit. The selective area grown NWFET (D1) is kept at 7.5 K and the 100 M$\Omega$ load resistor at room temperature. \textbf{b} Amplifier output, $V_\text{out}$,  as a function of  input, $V_\text{in}$, and bias voltage, $V_\text{DD}$. \textbf{c} Lorentzian fits of the derivative of the curves in panel (b). \textbf{d} Lorentzian's amplitude, $A$ (amplification), and half width, $\gamma$, as a function of bias. \textbf{e} Normalized frequency response of different NWFETs to a sinusoidal $V_\mathrm{G}$ of frequency $f$, measured at $1.2 \, \mathrm K$. 
    } 
    \label{Fig3}
  \hfill
\end{figure}

\indent Having established the overall characteristics of the InAs selective area grown NWFETs, we now consider their performance in the common source amplifier configuration shown in Fig.\ \ref{Fig3}a, with a NWFET operating at $7.5 \, \mathrm K$. Figure \ref{Fig3}b shows the amplifier output voltage, $V_\text{out}$, for various bias levels, $V_\text{DD}$, as a function of the input voltage, $V_\text{in}$, applied to the gate. The circuit operates as an inverting amplifier, as a positive change in $V_\text{in}$ results in a negative change in $V_\text{out}$. The amplification $A$ is determined as the peak amplitude in d$V_\text{out}$/d$V_\text{in}$, which was fitted to a Lorentzian line shape as shown in Fig.\ \ref{Fig3}c. The amplitudes and characteristic peak widths, $2\gamma=\text{FWHM}$, are shown in Fig.\ \ref{Fig3}d as a function of $V_\mathrm{DD}$. The maximum voltage gain $\sim 36$ is obtained at the highest $V_\mathrm{DD}$ and optimal gate-operation point $V_\text{0}\approx-0.86$ V. Upon shifting the operating point from $V_\mathrm 0$ by $\gamma$ the gain is reduced by a factor 2 and at high $V_\mathrm{DD}$ we find $\gamma \sim 10 \, \mathrm{mV}$ which is significantly smaller than the typical hysteresis $\Delta V_\text{G} \sim 50 \, \mathrm{mV}$ (Fig.\ 2f). This shows that for sub-threshold applications, such as amplifiers, the NWFET hysteresis should be reduced. Taking steps to improve the oxide and oxide/nanowire interface quality is likely to yield devices suitable for such applications. \\
\indent Due to the high mobility, InAs is often used in high bandwidth applications\cite{Blekker2010high, lind2024iii} and  Fig.\  3e shows the normalized current response of isolated NWFETs subject to a sinusoidal gate modulation of frequency $f$ (see supplementary section S6 and S7). For all devices, a cutoff frequency $\sim 5\, \mathrm{kHz}$, is found, matching the $5 \, \mathrm{kHz}$ cutoff of the cryostat filtering. This shows that, despite the relatively low mobility and significant interface charge trapping, the cutoff frequency of the NWFET exceeds this limit. Further measurements performed in a cryogenic high-bandwidth environment are needed to establish the upper operational limit, and additional optimization of SAG parameters, gate-oxide quality, and device design will likely be required to extend this limit.\\
\indent In summary, we have reported the fabrication and electrical characterization at cryogenic temperature of FETs based on InAs nanowires realized by  SAG.  The devices act as $n$-type depletion mode FETs, with $I_\mathrm{ON}/I_\mathrm{OFF}$ ratio exceeding 10$^3$, DIBL between 0.5 and 1.7 V/V and mobility ranging from 60 to 120 cm$^2$/Vs. The current-voltage characteristics exhibit hysteresis during gate sweeps and sub-threshold analysis reveals a density of interfacial traps on the order of 10$^{14}$ cm$^{-2}$ eV$^{-1}$.  The modest transistor characteristics could be improved by optimizing the nanowire growth to reduce misfit dislocations and material intermixing\cite{beznasyuk:2022}, while the hysteresis could be reduced by improving the oxide/nanowire interface quality \cite{d2023time}. Additional improvements should also focus on the device design to mitigate short-channel effects \cite{delalamo:2011}. Furthermore, we tested the NWFETs in a common source amplifier configuration exhibiting a maximum amplification of 36 at $V_\text{DD}=1$ V, and we observed a NWFET bandwidth higher than that of the cryogenic setup. Overall, the NWFETs operate effectively in saturation-mode applications, while the hysteresis remains a limitation for sub-threshold operations. Taken all together, however, the results demonstrate the potential of InAs selective area grown NWFETs as a scalable platform for cryogenic nanoelectronics.\\
\\
See the supplementary material for detailed information on the performed measurements, additional transfer characteristics data, the methodology used to extract threshold voltage and mobility, details on hysteresis computations and on bandwidth measurements in both the amplifier and NWFET configurations.
\section*{ACKNOWLEDGMENTS}
This work was supported by the European Research Council under the European Union’s Horizon 2020 research and innovation program (Grant no.: 866158). ICN2 acknowledges funding from Generalitat de Catalunya 2021SGR00457. Authors acknowledge the Advanced Materials programme by the Spanish Government with funding from European Union NextGenerationEU (PRTR-C17.I1) and by Generalitat de Catalunya (In-CAEM Project). We acknowledge support from CSIC Interdisciplinary Thematic Platform (PTI+) on Quantum Technologies (PTI-QTEP+). This research work has been funded by the European Commission – NextGenerationEU (Regulation EU 2020/2094), through CSIC's Quantum Technologies Platform (QTEP). ICN2 is supported by the Severo Ochoa program from Spanish MCIN / AEI (Grant No.: CEX2021-001214-S) and is funded by the CERCA Programme / Generalitat de Catalunya. Authors acknowledge the use of instrumentation as well as the technical advice provided by the Joint Electron Microscopy Center at ALBA (JEMCA). ICN2 acknowledges funding from Grant IU16-014206 (METCAM-FIB) funded by the European Union through the European Regional Development Fund (ERDF), with the support of the Ministry of Research and Universities, Generalitat de Catalunya. ICN2 is founding member of e-DREAM.

\section*{AUTHOR DECLARATIONS}
\subsection*{Conflict of Interest}
The authors have no conflicts to disclose.
\subsection*{Author Contributions}
\textbf{Giulia Meucci:} Formal analysis; Methodology; Visualization; Writing – original draft; Writing – review \& editing. 
\textbf{Dags Ol{\v{s}}teins:} Conceptualization; Data curation; Investigation; Writing – review \& editing.
\textbf{Damon J. Carrad:} Conceptualization; Data curation; Investigation; Writing – review \& editing.
\textbf{Gunjan Nagda:} Data curation; Investigation; Writing – review \& editing.
\textbf{Daria V. Beznasyuk:} Data curation; Investigation; Writing – review \& editing.
\textbf{Christian E. N. Petersen:} Methodology; Writing – review \& editing.
\textbf{Sara Martí-Sánchez:} Data curation; Investigation.
\textbf{Jordi Arbiol:} Data curation; Investigation.
\textbf{Thomas Sand Jespersen:} Conceptualization; Supervision; Project administration; Writing – original draft; Writing – review \& editing.
\section*{DATA AVAILABILITY}
The data that support the findings of this study are openly available at https://doi.org/10.11583/DTU.29859029.

\bibliography{ref,TSJZotero}

\end{document}



\author{Giulia Meucci}
\affiliation{Department of Energy Conversion and Storage, Technical University of Denmark, 2800 Kgs.Lyngby, Denmark}

\author{Dags Ol{\v{s}}teins}
\affiliation{Department of Energy Conversion and Storage, Technical University of Denmark, 2800 Kgs.Lyngby, Denmark}

\author{Damon J. Carrad}
\affiliation{Department of Energy Conversion and Storage, Technical University of Denmark, 2800 Kgs.Lyngby, Denmark}

\author{Gunjan Nagda}
\affiliation{Department of Energy Conversion and Storage, Technical University of Denmark, 2800 Kgs.Lyngby, Denmark}

\author{Daria V. Beznasyuk}
\affiliation{Department of Energy Conversion and Storage, Technical University of Denmark, 2800 Kgs.Lyngby, Denmark}

\author{Christian E. N. Petersen}
\affiliation{Department of Energy Conversion and Storage, Technical University of Denmark, 2800 Kgs.Lyngby, Denmark}

\author{Sara Martí-Sánchez}
\affiliation{Catalan Institute of Nanoscience and Nanotechnology (ICN2), CSIC and BIST, Campus UAB, \\ Bellaterra,  Barcelona, Catalonia, Spain}

\author{Jordi Arbiol}
\affiliation{Catalan Institute of Nanoscience and Nanotechnology (ICN2), CSIC and BIST, Campus UAB, \\ Bellaterra,  Barcelona, Catalonia, Spain}\affiliation{ICREA, Passeig de Lluís Companys 23, 08010 Barcelona, Catalonia, Spain}

\author{Thomas Sand Jespersen}
\affiliation{Department of Energy Conversion and Storage, Technical University of Denmark, 2800 Kgs.Lyngby, Denmark}\affiliation{Center For Quantum Devices, Niels Bohr Institute, University of Copenhagen, 2100 Copenhagen, Denmark}

\title{Supplementary Material for Cryogenic performance of field-effect transistors and amplifiers based on selective area grown InAs nanowires}

\maketitle


\renewcommand{\thesection}{S\arabic{section}}
\setcounter{figure}{0}
\renewcommand{\thefigure}{S\arabic{figure}}
\setcounter{equation}{0}
\renewcommand{\theequation}{S\arabic{equation}}
\onecolumngrid

\section{Details of performed measurements} \label{Details of performed measurements}
\raggedright Eight nominally identical devices (D1-D8) were measured, with each device having some or all of the following types of characterizations: output characteristics, transfer characteristics,  amplifier characteristics, amplifier bandwidth, and transistor bandwidth. Measurements were performed at  cryogenic temperatures and  voltage ranges indicated in the table below.
\begin{table}[h]

  \centering

   \begin{tabular*}{\textwidth}{@{\extracolsep{\fill}} c c c c c c @{}}
    \hline
    \multirow{3}{*}{Device} & \textbf{Output} & \textbf{Transfer} & \textbf{Amplifier} & \textbf{Amplifier} & \textbf{Transistor} \\
    & \textbf{characteristics} & \textbf{characteristics} & \textbf{characteristics} & \textbf{Bandwidth} & \textbf{Bandwidth} \\
    & \textbf{$I_\text{SD}$ vs $V_\text{SD}$ } & \textbf{$I_\text{SD}$ vs $V_\text{G}$} & \textbf{$V_\text{out}$ vs $V_\text{in}$} & \textbf{$A$ vs $f$} &\textbf{$I_\text{normalized}$ vs $f$} \\
    
    \hline
    
  D1 & Fig.\ 1b$^\text{a}$ & Fig.\ \ref{FigS2}$^\text{b}$ & Fig.\ 3b$^\text{f}$  & Fig.\ \ref{FigS4}$^\text{g}$  & \\

    D2 &   & Fig.\ \ref{FigS1}$^\text{c}$ &  & &  Fig.\ 3e $^\text{h}$ \\

    D3 &  &    Fig.\ \ref{FigS1}$^\text{c}$ &  &  & Fig.\ 3e, \ref{FigS3}$^\text{i}$ \\

    D4  & & Fig.\ 2a,b$^\text{c}$ & &  &  \\
 
    D5 & &  Fig.\ \ref{FigS1}$^\text{d}$ & & &  Fig.\ 3e$^\text{j}$ \\

    D6 &  &   Fig.\ \ref{FigS1}$^\text{d}$ &   && Fig.\ 3e$^\text{k}$ \\

    D7 &  & Fig.\ \ref{FigS1}$^\text{d}$ & &&  Fig.\ 3e$^\text{l}$ \\
    
    D8 &  & Fig.\ \ref{FigS1}$^\text{e}$ &  & &  \\
    
    \hline
   \multicolumn{6}{l}{ $^\text{a}$ $T=9$ K, $V_\text{SD}=0$ to $1$ V, $V_\text{G}=-0.9$ to $1$ V (steps $0.1$ V).}\\
   \multicolumn{6}{l}{ $^\text{b}$ $T=8.5$ K, $V_\text{G}=-1$ to $1$ V (sweep rate $\sim  60 \, \mathrm{mV}/\mathrm s$), $V_\text{SD}=0.1$ to $1$ V (steps $0.1$ V).}\\
   \multicolumn{6}{l}{ $^\text{c}$ $T=14$ mK, $V_\text{G}=-1$ to $1$ V (sweep rate $\sim  60 \, \mathrm{mV}/\mathrm s$), $V_\text{SD}=20$ to $100$ mV (steps $20$ mV).}\\
   \multicolumn{6}{l}{ $^\text{d}$ $T=14$ mK, $V_\text{G}=-1$ to $1$ V (sweep rate $\sim  60 \, \mathrm{mV}/\mathrm s$), $V_\text{SD}=100$ mV.}\\
   \multicolumn{6}{l}{ $^\text{e}$ $T=14$ mK, $V_\text{G}=-1$ to $0$ V (sweep rate $\sim  60 \, \mathrm{mV}/\mathrm s$), $V_\text{SD}=20$ to $100$ mV (steps $20$ mV).}\\
   \multicolumn{6}{l}{ $^\text{f}$ $T=7.5$ K, $V_\text{in}=-1$ to $-0.6$ V, $V_\text{DD}=0.1$ to $1$ V (steps $0.1$ V).}\\
   \multicolumn{6}{l}{ $^\text{g}$  $T=7.5$ K, $f=1$ to $10^4$ Hz, $V_\text{DD}=1$ V, $V_\text{0}=-0.882$ V.}\\
   \multicolumn{6}{l}{ $^\text{h}$  $T=1.2$ K, $f=10^2$ to $2 \cdot 10^5$ Hz, $V_\text{SD}=0.02$ V, $V_\text{G0}=-0.5$ V.}\\
   \multicolumn{6}{l}{ $^\text{i}$  $T=1.2$ K, $f=1$ to $2 \cdot 10^5$ Hz, $V_\text{SD}=0.1$ V, $V_\text{G0}=-0.6$ V.}\\
   \multicolumn{6}{l}{ $^\text{j}$  $T=1.2$ K, $f=10^2$ to $2 \cdot 10^5$ Hz, $V_\text{SD}=0.1$ V, $V_\text{G0}=-0.725$ V.}\\
   \multicolumn{6}{l}{ $^\text{k}$  $T=1.2$ K, $f=10^2$ to $2 \cdot 10^5$ Hz, $V_\text{SD}=0.1$ V, $V_\text{G0}=-0.65$ V.}\\
   \multicolumn{6}{l}{ $^\text{l}$  $T=1.2$ K, $f=10^2$ to $2 \cdot 10^5$ Hz, $V_\text{SD}=0.1$ V, $V_\text{G0}=0$ V.}
    
  \end{tabular*}
  \label{tab:simple}
\end{table}

\section{Transfer characteristics} \label{Transfer characteristics at 14mK}
\begin{figure}[H]
  \centering
    \includegraphics[width = \linewidth]{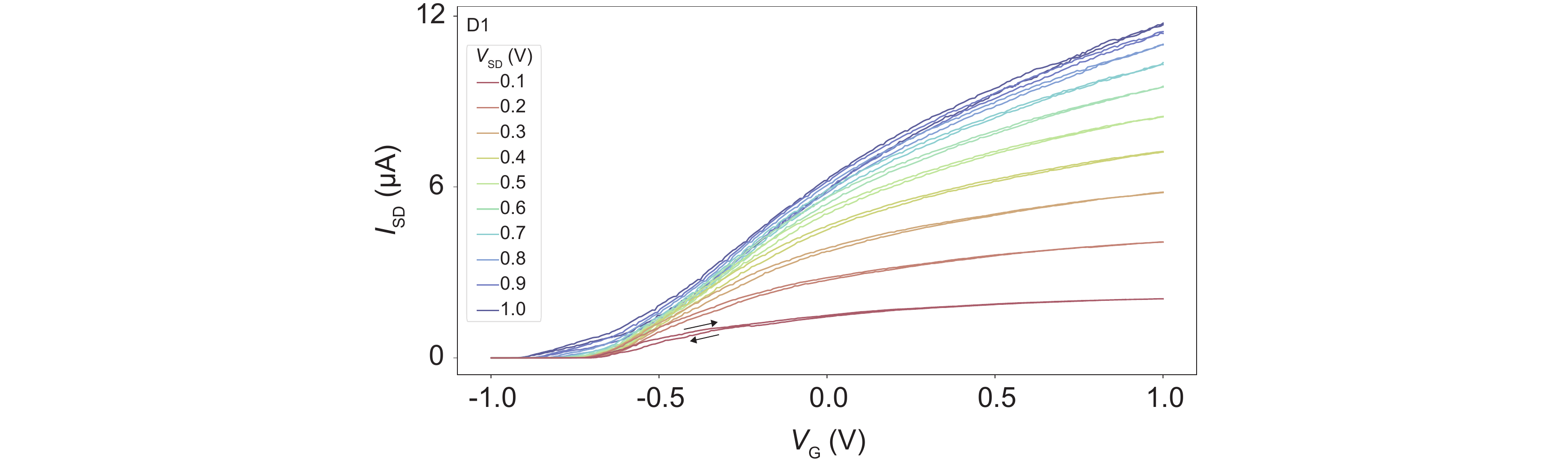}
    \caption{Transfer characteristics of D1 measured at 7.5 K, for different values of $V_\text{SD}$, from $0.1$ V (red) to $1$ V (blue).
    } 
    \label{FigS2}
\end{figure}

\begin{figure}[H]
  \centering
    \includegraphics[width = \linewidth]{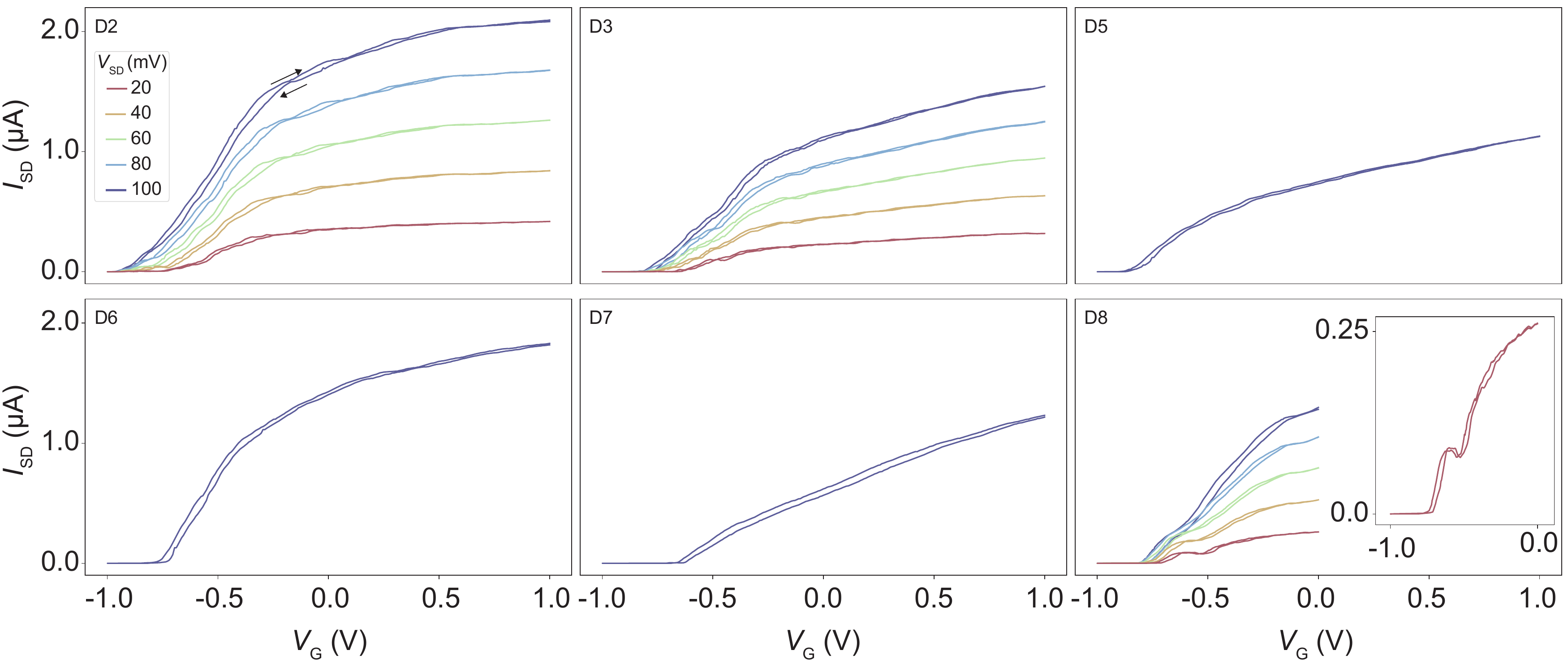}
    \caption{Transfer characteristics of several devices measured at 14 mK, for different values of $V_\text{SD}$, ranging from $20$ mV (red) to $100$ mV (blue). D8 shows effects of quantum interference and ballistic transport at low bias. 
    } 
    \label{FigS1}
\end{figure}

The transfer characteristics ($I_\text{SD}$ vs $V_\text{G}$) of all eight devices were measured for different $V_\text{SD}$ values. For D1, measurements were conducted at 8.5 K, with results shown in Fig.\ \ref{FigS2}. For the remaining devices, measurements were conducted at 14 mK. Results for D4 are presented in Fig.\ 2a, while results for the other devices are shown in Fig.\ \ref{FigS1}. D8 exhibited conductance resonances and plateaus likely due to quantum interference effects, possibly coupled with ballistic transport modes (Fig.\ \ref{FigS1}). Such behaviours are not captured by traditional FET models and thus D8 was excluded from analysis.  

\section{Extracting threshold voltage and mobility} \label{extractingthreshold}

\begin{figure}[H]
  \centering
    \includegraphics[width = \linewidth]{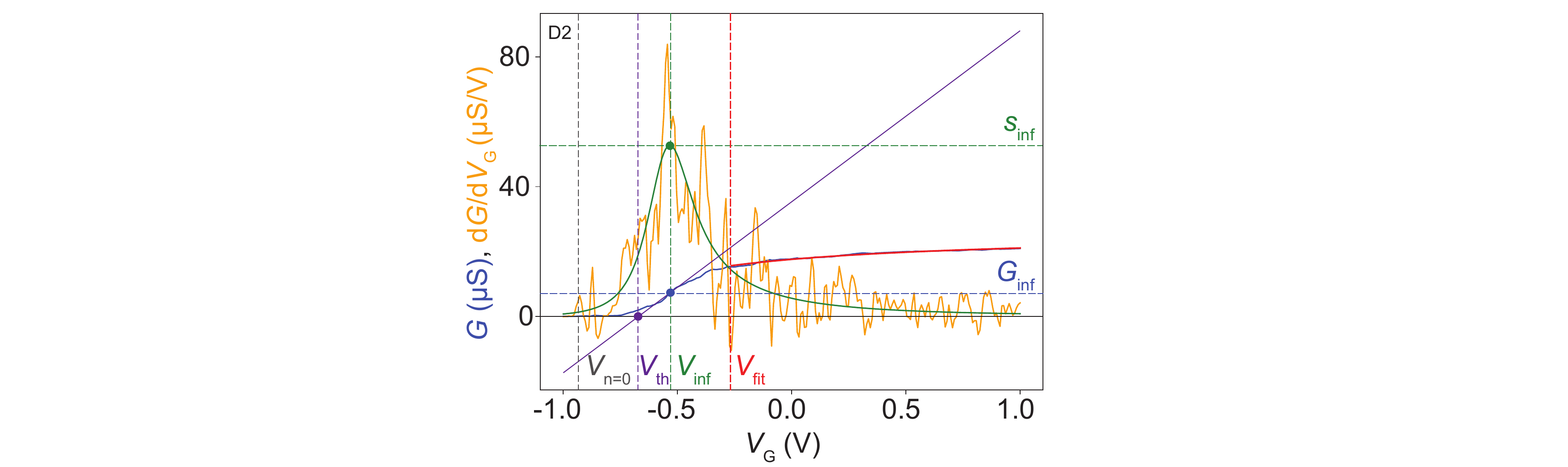}
    \caption{ Extracting the threshold voltage and mobility for D2 following the procedure outlined in Ref.\ \citenum{gatedependentmobility}. The conductance (blue line) is used to compute the trans-conductance (yellow line), which is fitted by an asymmetric Lorentzian (green line). The coordinates of the Lorentzian peak (green dot), $V_\text{inf}$ and $s_\text{inf}$ (green dashed lines), are used to compute the tangent to the conductance curve at inflection point (purple line). The threshold voltage, $V_\text{th}$, is defined as the point where the tangent line is zero (purple dot and dashed line) and the zero density voltage is defined as $V_\text{n=0}=V_\text{th}-2|V_\text{inf}-V_\text{th}|$ (grey dashed line). The conductance is fitted to the Drude expression (red line), with the fit starting at $V_\text{fit}=V_\text{inf}+2|V_\text{inf}-V_\text{th}|$ (red dashed line), to extract the series resistance, which is then used to compute the gate-dependent mobility shown in the main text, Fig.\ 2f.  
    } 
    \label{FigSmobility}
\end{figure}

The threshold voltage and mobility were calculated following the procedure outlined in Ref.\ \citenum{gatedependentmobility}. First, the transfer curves, $I_\text{SD}$ vs $V_\text{G}$, are normalized by the applied $V_\text{SD}$, yielding the conductance $G(V_\text{G})$  (blue line in Fig.\ \ref{FigSmobility}). The trans-conductance, $\text{d}  G/\text{d} V_\text{G}$, is computed (yellow line in Fig.\ \ref{FigSmobility}) and fitted with an asymmetric Lorentzian:
\begin{equation}
   \frac{\text{d}  G}{\text{d} V_\text{G}} =\frac{2  A}{\pi  U_\text{0}}  \left(1 + 4  \left(\frac{V_\text{G} - V_\text{G0}} {U_\text{0}}\right)^2\right)^{-1} \, , 
\label{eq.al}
\end{equation}
where 
\begin{equation}
    U_\text{0}= 2c \left(1+\exp(a(V_\text{G} - V_\text{G0}))\right)^{-1} \, , 
\end{equation}
and $A$, $V_\text{G0}$, $c$ and $a$ are fitting parameters (green line in Fig.\ \ref{FigSmobility}). The coordinates of the Lorentzian peak are the gate voltage, $V_\text{inf}$, and the trans-conductance, $s_\text{inf}=\text{d}  G/\text{d} V_\text{G}|_{V_\text{G}=V_\text{inf}}$, at the inflection point (green dots and dashed lines in Fig.\ \ref{FigSmobility}). With this information, the tangent line at the inflection point is:
\begin{equation}
   T(G)=G_\text{inf}+s_\text{inf}(V_\text{G}-V_\text{inf}) \, , 
\end{equation}
(purple line in Fig.\ \ref{FigSmobility}) where $G_\text{inf}=G(V_\text{inf})$  (blue dot and dashed line in Fig.\ \ref{FigSmobility}). The threshold voltage, $V_\text{th}$ is, then, defined as the point where $T(G)=0$ (purple dot and dashed line in Fig.\ \ref{FigSmobility}), and the zero density voltage is defined as $V_\text{n=0}=V_\text{th}-2|V_\text{inf}-V_\text{th}|$ (grey dashed line in Fig.\ \ref{FigSmobility}). 

Next, the series resistance, $R_\text{s}$,  is estimated fitting the conductance to the Drude expression: 
\begin{equation}
  G  = \left(R_\text{s, fit}+\frac{L^2}{\mu_\text{fit} C (V_\text{G}-V_\text{n=0} )} \right)^{-1} \, , 
\end{equation}
where $R_\text{s, fit}$ and $\mu_\text{fit}$ are fitting parameters and $L$ is the device length (red line in Fig.\ \ref{FigSmobility}). In this step the starting point for the fit is fixed at  $V_\text{fit}=V_\text{inf}+2|V_\text{inf}-V_\text{th}|$ (red dashed line in Fig.\ \ref{FigSmobility}). 

Finally, the series resistance is subtracted from the total resistance to obtain the device conductance and extract the gate-dependent mobility, $\mu$:
\begin{equation}
  \mu (V_\text{G})  = \frac{L^2}{ C (V_\text{G}-V_\text{n=0} )(\frac{1}{G(V_\text{G})}-R_\text{s, fit})}  \, .
\end{equation}

\section{Hysteresis computations} \label{Hysteresis computations}
\begin{figure}[H]
  \centering
    \includegraphics[width = \linewidth]{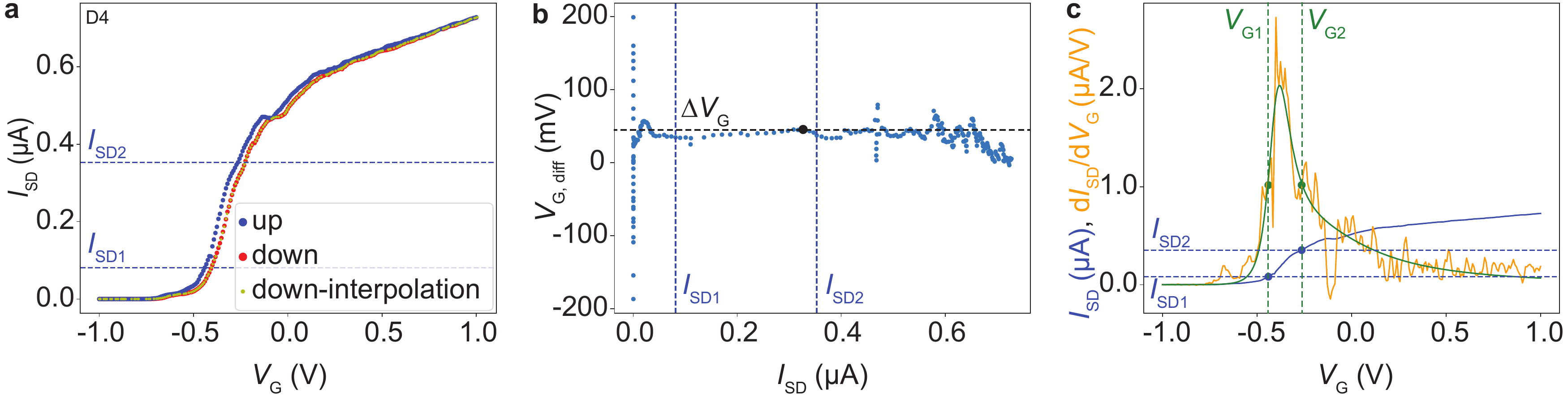}
    \caption{\textbf{a} Transfer characteristics of D4 measured at $V_\text{SD}=40$ mV and $T=14$ mK, during an up (blue) and down (red) sweep. The down-sweep measurement points are interpolated to match the current of the up-sweep points (yellow). The blue dashed lines, $I_\text{SD1}$ and $I_\text{SD2}$, represent the range in which $\Delta V_\text{G}$ is evaluated. \textbf{b} Voltage difference between the up and down sweep, $V_\text{G, diff}$,  as a function of $I_\text{SD}$. $\Delta V_\text{G}$ is defined as the maximum $V_\text{G, diff}$ in the range limited by the blue dashed lines, $I_\text{SD1}$ and $I_\text{SD2}$.  \textbf{c} Derivative (orange line) of the up sweep transfer characteristics (blue line), fitted by an asymmetric Lorentzian (green line).  $V_\text{G1}$ and $V_\text{G2}$ (green points and green dashed lines) define the full-width half-maximum. On the transfer characteristics, these corresponds to $I_\text{SD1}$ and $I_\text{SD2}$ (blue points and blue dashed lines), which determine the range in which $\Delta V_\text{G}$ is evaluated. 
    } 
    \label{FigS5}
\end{figure}

\raggedright To evaluate the hysteresis we measure $I_\text{SD}$ while sweeping $V_\text{G}$ up ($I_\text{SD, up}$) and down ($I_\text{SD, down}$). We are interested in the $V_\text{G}$ difference between the two sweeps as a function of  $I_\text{SD}$. To compare the two curves, it is necessary to interpolate one of them so that both curves share the same $I_\text{SD}$ values. We define $I_\text{SD}=I_\text{SD, up}$ so that $V_\text{G, up}=V_\text{G}$, and we interpolate the down-sweep measurement points to obtain $V_\text{G, down}$ as a function of $I_\text{SD}$ (see Fig.\ \ref{FigS5}a).  Figure \ref{FigS5}b shows $V_\text{G, diff}=(V_\text{G, down}-V_\text{G, up})$ for D4 and $V_\text{SD}=40$ mV. This difference is nearly constant except at extreme $I_\text{SD}$ where the transfer curves saturate. Therefore, it appears necessary to limit the evaluation of $V_\text{G, diff}$ to the region close to the inflection point of the transfer curve. This region is defined by fitting the derivative of the up-sweep transfer curve to an asymmetric Lorentzian, of the form described in equation \ref{eq.al} (Fig.\ \ref{FigS5}c). We determine the two points on the asymmetric Lorentzian corresponding to half the peak value, $V_\text{G1}$ and $V_\text{G2}$, which define FWHM$_\text{AL}=|V_\text{G2}-V_\text{G1}|$. The current values, $I_\text{SD1}$ and $I_\text{SD2}$, corresponding to these voltages   are used to delimit the range in which  $V_\text{G, diff}$ is evaluated. The hysteresis parameter $\Delta V_\text{G}$, discussed in the manuscript (Fig.\ 2e) is then defined as the maximum $V_\text{G, diff}$ in this range (Fig.\ \ref{FigS5}b). 

\section{Bandwidth measurements in the amplifier configuration} \label{Bandwidth measurement in the amplifier configuration}
\begin{figure}[H]
  \centering
    \includegraphics[width = \linewidth]{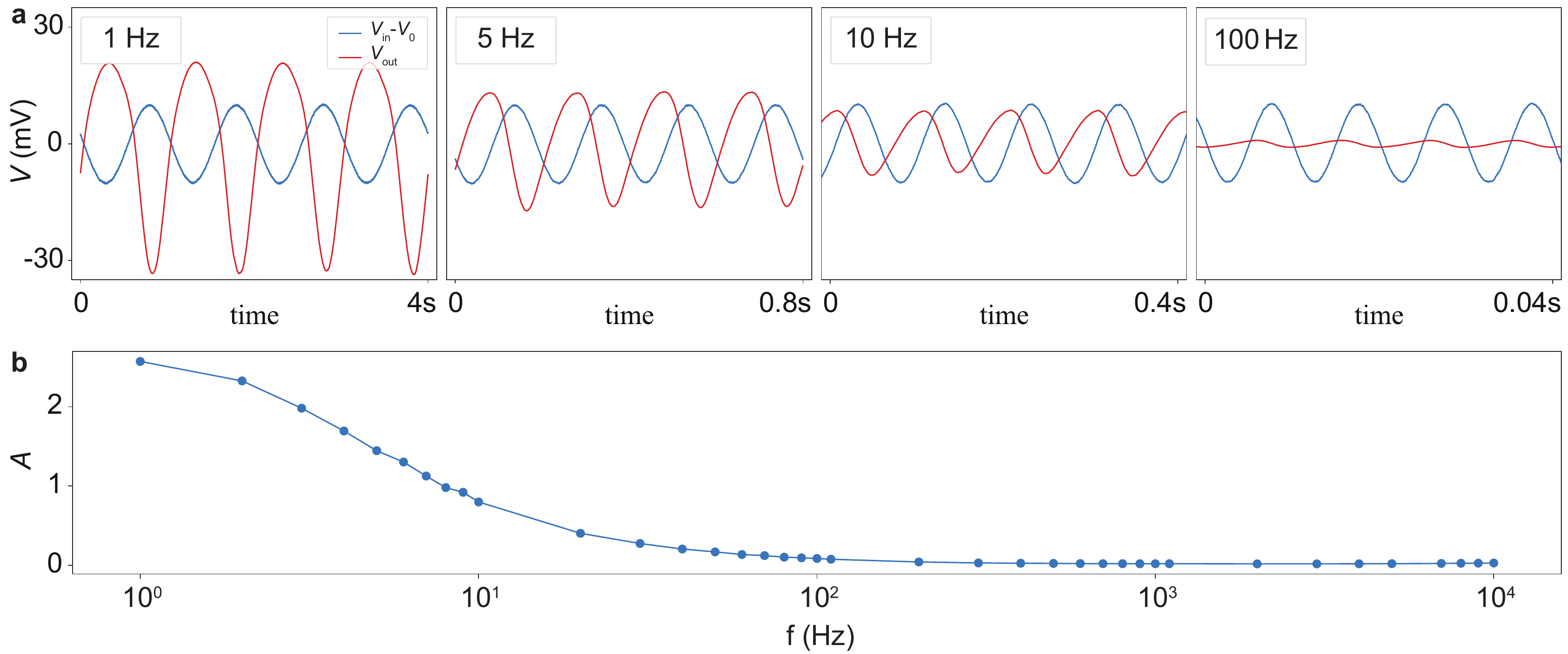}
    \caption{\textbf{a} $V_\text{out}$ response (red) to a sinusoidal $V_\text{in}$ (blue) of different frequencies, for the amplifier configuration shown in Fig.\ 3a . \textbf{b} Frequency dependence of the gain, computed as the ratio of the amplitudes of the sinusoidal fit of  $V_\text{out}$ and $V_\text{in}$. The measurements are performed at  at 7.5 K. 
    } 
    \label{FigS4}
\end{figure}

\raggedright To investigate the device bandwidth in the amplifier configuration with D1 shown in Fig.\ 3a we applied a sinusoidal input voltage at varying frequencies $f$:
\begin{equation}
    V_\text{in}=V_\text{0}+A_\text{in} \sin(2\pi f t+\phi_\text{in}) \, ,
\end{equation}
where $V_\text{0}=-0.882$ V is the optimal gate-operation point of the amplifier, $A_\text{in}=0.01$V is the amplitude, $t$ is the time and $\phi_\text{in}$ is the phase. We fitted the measured $V_\text{out}$  to a sine wave of the form:
\begin{equation}
    V_\text{out}=A_\text{out}\sin(2\pi f t+\phi_\text{out}) \, ,
\end{equation}
where $A_\text{out}$ and $\phi_\text{out}$ are fitting parameters (Fig.\ \ref{FigS4}a).  The gain, $A$, was calculated as $A=A_\text{out}/A_\text{in}$. Figure \ref{FigS4}b shows that $A$ decreases with increasing $f$, starting from frequencies below 1 Hz, which is much lower than the RC filter cutoff frequency $f_\text{c}=5$ kHz. This is due to the load resistance, $R_\text{load}=100$ M$\Omega$, which effectively reduce the cutoff frequency:
\begin{equation}
    f_\text{c, amp}=\frac{1}{2\pi R C_f}=\frac{1}{2\pi (R_\text{f}+R_\text{load}) C_f}=f_\text{c}\frac{R_\text{f}}{R_\text{f}+R_\text{load}}\approx0.3 \text{Hz}\, ,
\end{equation}
where $R_\text{f}\approx6$ k$\Omega$ and $C_f$ are the resistance and capacitance of the cryostat filters. Consequently, it is impossible to measure bandwidths higher than $\approx0.3 \text{Hz}$ in the amplifier configuration, and, thus, the bandwidth is analyzed by studying the frequency response of the isolated NWFET.

\section{NWFET bandwidth measurements} \label{Bandwidth measurements}
\begin{figure}[H]
  \centering
    \includegraphics[width = \linewidth]{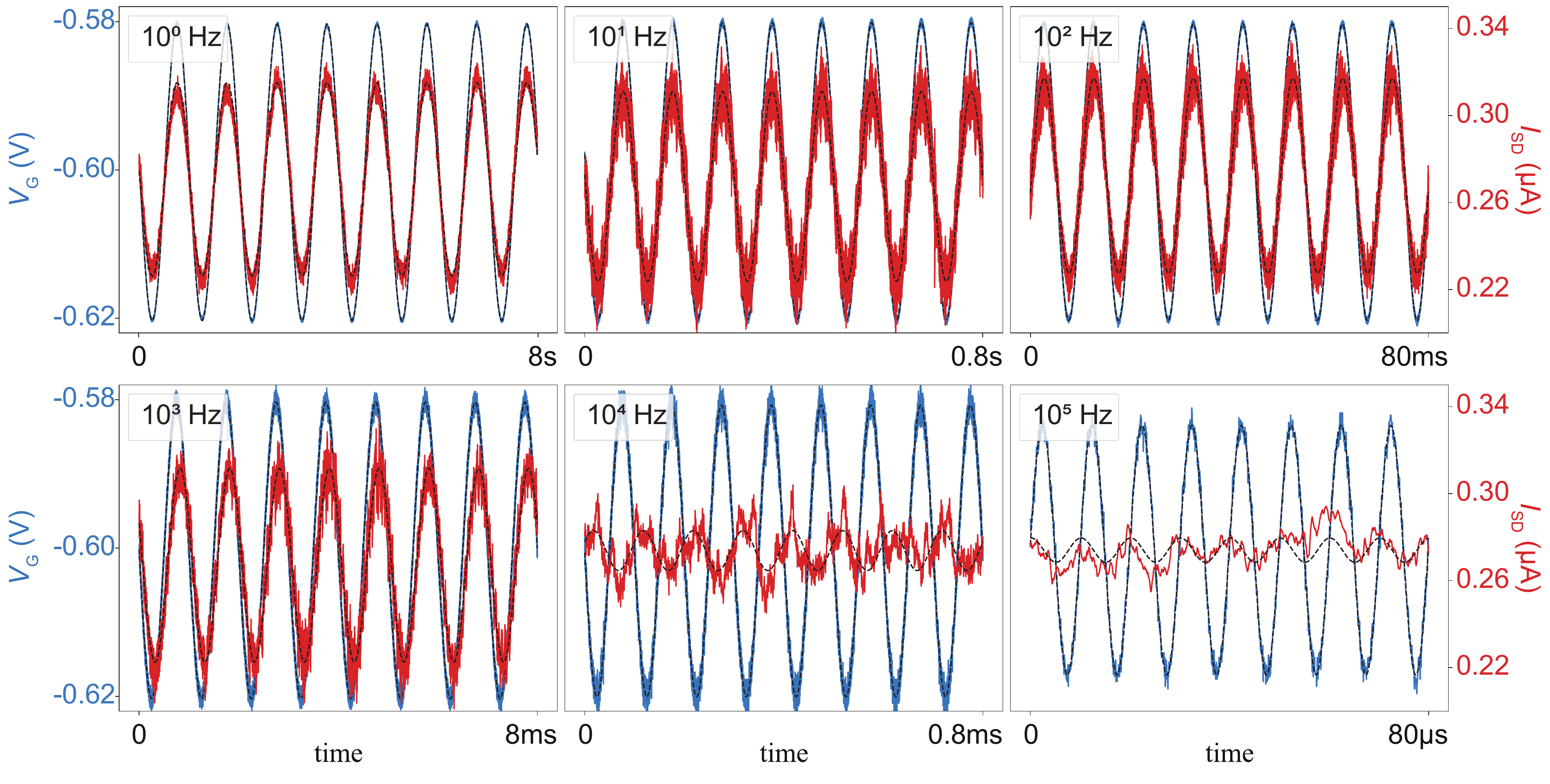}
    \caption{Current response (red) to a sinusoidal gate voltage (blue) of different frequencies for device D3 at 1.2 K. The black dashed lines indicate the sinusoidal fits. 
    } 
    \label{FigS3}
\end{figure}

Similarly to what discussed in supplementary section S5, the bandwidth of the NWFETs is measured by applying a sinusoidal gate voltage:
\begin{equation}
    V_\text{G}=V_\text{G0}+A_{V_\text{G}}\sin(2\pi f t+\phi_{V_\text{G}}) \, ,
\end{equation}
where $A_{V_\text{G}}=20$ mV and $V_\text{G0}$ varies depending on the device as described supplementary section \ref{Details of performed measurements}. The measured  $I_\text{SD}$ is fitted to a sine wave of the form:
\begin{equation}
    I_\text{SD}=A_{I_\text{SD}}\sin(2\pi f t+\phi_{I_\text{SD}}) \, ,
\end{equation}
as shown Fig.\ \ref{FigS3}. If the fit has a coefficient of determination, $R^2$, lower than 0.25, the dataset is excluded from the analysis. 
The frequency response is computed by normalizing the value of $A_{I_\text{SD}}$ by the value measured at the lowest frequency ($f_\text{min}=$ 1 or 100 Hz):
\begin{equation}
    I_\text{normalized}(f)=\frac{A_{I_\text{SD}}(f)}{A_{I_\text{SD}}(f=f_\text{min})} \, .
\end{equation}
Figure 3e shows $I_\text{normalized}$ as a function of $f$ for various devices.  The final signal at high frequency could be due to capacitive cross-talk of the cryostat wiring. 

\bibliography{ref,TSJZotero}